\documentstyle[prc,aps]{revtex}      
\input epsf
\epsfverbosetrue
\begin{document} 
\draft
\title{ Uncertainties in the 
$\eta$-Nucleon Scattering Length and Effective Range}

\author{A.M.~Green\thanks{email:  green@pcu.helsinki.fi}}
\address{Department of Physics, University of Helsinki and Helsinki Institute 
of Physics, P.O. Box 9, FIN--00014, Finland}
\author{S.~Wycech\thanks{email:  wycech@fuw.edu.pl}}
\address{Soltan Institute for Nuclear Studies,Warsaw, Poland}
\date{\today}
\maketitle
\begin{abstract}  
The coupled  $\eta N$ , $\pi N$, $\gamma N$ system is described by a K-matrix  
method. The parameters in this model are adjusted to get an optimal
fit to $\pi  N\rightarrow \pi  N$, $\pi  N\rightarrow \eta N$,
$\gamma N\rightarrow \pi N$ and
$\gamma N\rightarrow \eta N$ data in an energy range of about 100MeV or
 so each side of the $\eta$ threshold. The outcome is the appearance
of two solutions one which has an $\eta$-nucleon scattering length ($a$)
of about 1.0 fm and a second with $a\approx 0.2$fm. 
However, the second
solution has an unconventional non-Lorentian form for the $T$-matrix 
in the region of the 1535(20)MeV and 1650(30)MeV  $S$-wave $\pi N$ 
resonances.\\

\end{abstract} 
 \pacs{PACS numbers: 13.75.-n, 25.80.-e, 25.40.V }

The value of the  $\eta$-nucleon scattering length ($a$) is still uncertain,
but with everyone agreeing that it is indeed attractive i.e. $a>0$.
In the literature estimates can be found ranging from  $Re \ a <0.3$fm
\cite{Speth} upto  about 1.0$\pm 0.1$fm \cite{gw99} -- a selection being
given in Table I. 

The main interest in
$a$ lies in the fact that, if $a$ is sufficiently attractive, then
$\eta$-nuclear quasi-bound states may be possible. These were first
suggested about 15 years ago \cite{hai,li}. Since then many articles
have appeared on this subject studying different reactions in which
such quasi-bound states could manifest themselves e.g. in the   
$\eta$-deuteron system \cite{ued},  
the $ pd \rightarrow \  ^{3}He \eta$ reaction \cite{wil,gar} and in
the  $ pp \rightarrow pp \eta$ cross section \cite{cal}.   
However,  the first experimental attempt to discover $\eta$-states in 
heavier nuclei gave a negative conclusion  \cite{chrien}. Another 
experiment is now  proposed  
to check this result \cite{hayano,tsu}.
Unfortunately, in the absence of $\eta$-beams,  these same reactions are
also the only source of experimental information about the $\eta N$
scattering length. It is, therefore, important that in any discussion
as many reactions as possible are treated simultaneously. Otherwise
success with one reaction may be completely nullified by failure with
another. 

With this in mind, in Ref.\cite{gw97}, the present authors 
carried out a
simultaneous $K$-matrix fit to the $\pi N\rightarrow\eta N$ cross
sections reviewed by Nefkens \cite{Nefkens} and the $\gamma p\rightarrow\eta p$
data of Krusche et al. \cite{Krusche}. In addition, the fit included   
$\pi N$ amplitudes of Arndt et al. \cite{Arndt}, since the $\pi N$ and 
$\eta N$ channels are so strongly coupled. Using the notation for the 
$T$-matrix: 
$T^{-1}+iq_{\eta}=1/a+\frac{r_0}{2} q^2_{\eta}+s q^4_{\eta}$,
 -- with $ q_{\eta}$ being the $\eta$-momentum in the $\eta N$ center-of-mass --
 resulted in the $\eta N$-effective range parameters

$a$(fm) = 0.75(4)+i0.27(3), $r_0$(fm) = --1.50(13)--i0.24(4)  and
$s$(fm$^3$) = --0.10(2)--i0.01(1) 

A later paper involving the present authors \cite{gw98} developed this 
formalism to look for complex poles in the $S$-matrix. In Ref.\cite{gw97}
the only place, where the $\eta N$ amplitude $ T_{\eta\eta}$ 
was directly checked against the data, 
was under the assumption that it was proportional to the photoproduction
amplitude -- namely --

\[ \tau(\gamma\eta)_r=
A(Phot) \sqrt{(Re \ T_{\eta\eta})^2+(Im \ T_{\eta\eta})^2},
 \ \ \  {\rm where} \ \  
\tau(\gamma\eta)_r=
\sqrt{\sigma(\gamma\eta)\frac{E_{\gamma}}{4\pi q_{\eta}}}\]
and $A(Phot)$ was a free parameter adjusted alongwith other parameters  to
ensure a good fit to the combined data. However, in Ref.~\cite{gw98} 
an alternative assumption, that $A(Phot)$ could also be  energy independent,
was also considered, namely, 
\begin{equation}
\label{A}
A(Phot)\propto \alpha(1+iT_{\eta \eta})+\frac{\beta}{q_{\eta}}T_{\eta \eta}
\end{equation}
-- a form  that had proven successful in Ref.\cite{Arndt2} when
describing pion photoproduction. This led to two distinct solutions
called GW11 and GW21. The former, with the energy independent $A(Phot)$,
resulted in parameter values  for the complex poles 
that were close to those given by the Particle Data Group \cite{PDG} and
are here referred to as conventional (C).
On the other hand the use of Eq.\ref{A} 
yielded very unconventional (UC) values. 
In particular, the coupling of  the $\pi N $ channels  to the
resonances at 1535(20)MeV and 1650(30)MeV generated widths of 354 and 133 MeV
respectively, whereas the GW11 solution gave 152 and 150 in agreement
with the PDG values of the resonance widths of  150(5) and 145 -- 190 
(all in MeV). 
Of course, it must be remembered that the PDG values are for 
poles in the $T$-matrix and not the $K$-matrix and that the nomenclature C
versus UC is 
based on the underlying assumption that the $T$-matrix and
$K$-matrix poles are closely related. Therefore, if, with an improved
data basis, the GW21 type of solution becomes the preferred solution --
not the case at present -- then it could indicate 
that the $K$-matrix  formulation  generates strongly energy dependent 
widths. 
These differences in the widths are also reflected in the  
effective range parameters ($a, \ r_0, \ s$). As expected,  GW11 gives values
for $a, \ r_0, \ s$ that are very similar to those of the earlier fit in
Ref.\cite{gw97}, whereas those from GW21 are very different\footnote{The
authors wish to thank Bengt  Karlsson for pointing this out.} being

\begin{figure}[ht]
\includegraphics{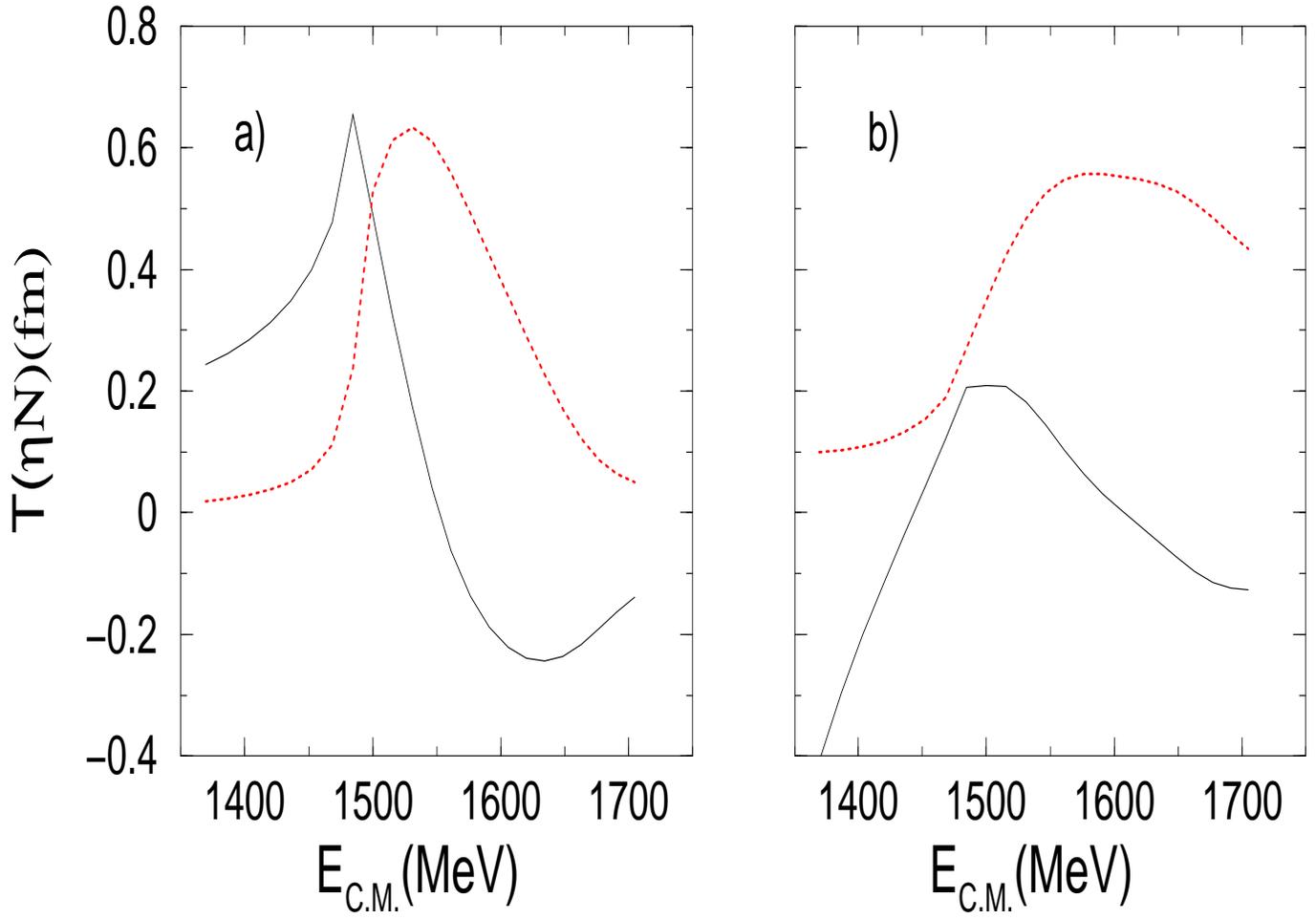}
\caption{The $\eta N\rightarrow \eta N$ amplitude  $T(\eta N)=T_{\eta
\eta}$ in fm as a function of the center-of-mass energy $E_{C.M.}$ in MeV.
Solid line for Re $T$ and Dotted line for Im $T$. 
a) is for the conventional(C) solution GW11 and 
b) is for the unconventional(UC) solution GW21  }
\label{fig1.}
\end{figure}

$a$(fm) = 0.21+i0.30, $r_0$(fm) = --2.61+i6.67  and
$s$(fm$^3$) = --0.39--i3.67, 

where the large values of the Imaginary parts give support to the possible
need for energy dependent widths.
This wild behaviour with such large imaginary components shows that the 
effective range expansion, in contrast to GW11, is poor for GW21.
The results from Ref.\cite{gw97} and GW11 are within the error bars
quoted in \cite{gw97}. However, they are not identical, since a slightly
different selection of data and their associated error bars was used.
In discussions concerning quasi-bound states, it is the value of
$T_{\eta \eta}$ in the unphysical region a few 10's of MeV below the 
$\eta N$-threshold that is relevant. 
In Figs.~\ref{fig1.} a) and b) this is shown
for both GW11 and GW21. There it is seen that  the GW21 amplitudes are 
not only smaller than those for
GW11 at the threshold but also they tend to drop faster over the range
below the threshold that is important for quasi-bound states.
The conclusion from \cite{gw97} is that the $\eta N$-effective range
parameters seem to be crucially dependent on the form of $A(phot)$ -- a
point that cannot be resolved within the restricted form of the 
$K$-matrix model presented there.

In a third paper by the authors \cite{gw99}, the $K$-matrix model was
extended so that, not only are the  $\pi N$ and $\eta N$ channels treated
explicitly -- as in Ref.\cite{gw97} --, but also the $\gamma N$ channel.
This means that the photoproduction reactions $\gamma N\rightarrow \pi N$
and $\gamma N\rightarrow \eta N$ can be treated on the same footing
as $\pi  N\rightarrow \pi  N$ and $\pi  N\rightarrow \eta N$. This
avoids the need to choose a form for $A(phot)$ to relate the $\gamma
N\rightarrow \eta N$ and $\eta N\rightarrow \eta N$ processes.
The penalty that is paid for this extension is that four more parameters are
needed. In the two channel form of Ref.\cite{gw97}, the basic parameters
[$E_{0,1} , \ \gamma_{\pi,\eta}(0,1), \  B_{ij}$] 
were those describing the various $K$-matrices -- namely --
\begin{equation}
 K_{\pi\pi}=\frac{\gamma_{\pi}(0)}{E_0-E}+
\frac{\gamma_{\pi}(1)}{E_1-E}
 \ \   , \ \  K_{\pi\eta}= 
\frac{\sqrt{\gamma_{\pi}(0)\gamma_{\eta}}}{E_0-E}+B_{\pi\eta}
 \ \  , \ \
 K_{\eta\eta}= \frac{\gamma_{\eta}}{E_0-E}+B_{\eta\eta}.
\end{equation}
The $E_{0,1}$ are the positions of poles that in a "conventional" model
should be near the energies of the $S$-wave $\pi N$ resonances $N(1535)$
and $N(1650)$. The $\gamma_{\pi,\eta}(0,1)$ are channel coupling
parameters that are related to the widths of these resonances. Again
these widths are thought to be more or less known when data is analysed by a
"conventional" model. However, as mentioned above, less conventional
models can lead to widths that are quite different. Finally the
$B_{ij}$ are energy independent background terms and are purely 
phenomenological. In the extended $K$-matrix model of  Ref.\cite{gw99}
the introduction of an explicit  $\gamma N$ channel requires the
additional $K$-matrices containing four more parameters
$\gamma_{\gamma}(0,1), \ B_{\gamma\eta}, \ B_{\gamma \pi}$:
\begin{equation}
\label{ketagamma}
K_{\gamma\eta}=\frac{\sqrt{\gamma_{\gamma}(0)\gamma_{\eta}}}
{E_0-E}+B_{\gamma\eta} \ \ , \ \ 
K_{\gamma \gamma}=\frac{\gamma_{\gamma}(0)}{E_0-E}+
\frac{\gamma_{\gamma}(1)}{E_1-E}
\ \ , \ \
K_{\gamma \pi}=
\frac{\sqrt{\gamma_{\gamma}(0)\gamma_{\pi}(0)}}{E_0-E}+
\frac{\sqrt{\gamma_{\gamma}(1)\gamma_{\pi}(1)}}{E_1-E}
+B_{\gamma \pi}.
\end{equation}
It should be added that the $\pi\pi N$ channel was also included
implicitly in the formalism at the expense of two more parameters 
$\gamma_3(0,1)$ that are related to the $\pi\pi$ branching of the two 
resonances. These lead to four more $K$-matrices:
\[ K_{33}= \frac{\gamma_3(0)}{E_0-E}+\frac{\gamma_3(1)}{E_1-E} \ ,
\ \ \ \ \  K_{\pi 3}= \frac{\sqrt{\gamma_{\pi}(0)\gamma_{3}(0)}}{E_0-E}
+\frac{\sqrt{\gamma_{\pi}(1)\gamma_{3}(1)}}{E_1-E} \ , \ \ \
 K_{\eta 3}= \frac{\sqrt{\gamma_{\eta}\gamma_{3}(0)}}{E_0-E},\]
\begin{equation}
\label{kgamma3}
K_{\gamma 3}=
\frac{\sqrt{\gamma_{\gamma}(0)\gamma_{3}(0)}}{E_0-E}+
\frac{\sqrt{\gamma_{\gamma}(1)\gamma_{3}(1)}}{E_1-E}.
\end{equation}
The $\pi\pi N$ channel is then removed by making, for example, the 
replacement
\begin{equation}
\label{kee}
K_{\eta\eta}\rightarrow  \frac{\gamma_{\eta}}
{E_0-E}+B_{\eta\eta}
+i\frac{K_{\eta 3}q_3K_{3 \eta}}{1-iq_3K_{33}},
\end{equation}
where $q_3$ is a three body phase space element.
It should be emphasised that this treatment of the $\pi\pi N$ channel is
not an approximation. It is purely for convenience.
More details can be found in Ref.~\cite{gw99}. The outcome of that work
was essentially two distinct solutions (A and B) depending on the number of 
$\gamma N\rightarrow \pi N$ data points taken from Ref.~\cite{gpivip}.
Using only 32 points with  $E_{cm}\le 1550$ MeV gave $a(fm)=0.87+i0.27$
for solution A, whereas with 48 data points and $E_{cm}\le 1650$ MeV gave
$a(fm)=1.05+i0.27$ for solution B -- values not very different to the 
$a(fm)=0.75(4)+i0.27(3)$ of our earlier work \cite{gw97}.

The two solutions given in Figs.~\ref{fig1.} a) and b)
 may be understood 
in terms of a simpler model. There are three  basic components 
of  $K_{\eta\eta}$ in Eq.~\ref{kee}.
The pole term with $E_0-E = \Delta $ is due to the $N(1535)$
and $\gamma_{\eta}$ is the appropriate coupling constant. The  
second term $B_{\eta\eta}$ represents  background interactions due 
to an unknown potential in the elastic channel. The final term,
which we now call $i K_{ine}$ 
 is a contribution from the inelastic (mostly $\pi N$) channels. 
It contributes the absorptive part to  $K_{\eta\eta}$ and this 
is stressed by the factor $i$.  Close to the threshold we can write 
$\Delta=E_0-M_{\eta}-M_{N}-E_{kin}= \Delta_{o}-q^2/2 m_{\eta N}$, which
induces some energy dependence and determines the effective range 
expansion. The scattering length now becomes 

\begin{equation}
\label{s2}
A_{\eta\eta}(E)= \gamma_{\eta}\frac{1}{\Delta_o}+ B_{\eta\eta} + i K_{ine},
\end{equation}
and the energy dependence of the $K$ matrix is given by 
\begin{equation}
\label{s3}
\frac{1}{K_{\eta\eta}(E)}=\frac{1}{A_{\eta\eta}(0) + \frac{q^2\gamma_{\eta}}
{2 \Delta_o  m_{\eta N}}},
\end{equation}
which reduces to the usual effective range expression 
\begin{equation}
\label{s3a}
\frac{1}{K_{\eta\eta}(E)}=\frac{1}{A_{\eta\eta}(0)} -\frac{q^2\gamma_{\eta}}
{ A_{\eta\eta}^2 \Delta_o^2 2 m_{\eta N}},
\end{equation}
provided $A_{\eta\eta}(0)$ is not small.
Now, the  solution of Fig.~\ref{fig1.}a) corresponds to a dominance of the
pole term 
over the background term $B_{\eta\eta}$. This induces a large scattering length 
and also a large effective range, so that  Eq.~\ref{s3a} is valid.
 Nevertheless, the alternative  
expansion in Eq.~\ref{s3} also converges very fast in a broad energy range --
see Fig.~2 in Ref.~\cite{gw97}. 
On the other hand, the solution of Fig.~\ref{fig1.}b) corresponds  to a strong 
cancellation of the pole  and the background terms
  $ \gamma_{\eta}\frac{1}{\Delta_o} + B_{\eta\eta}  \approx 0 $, 
so that $Re \ A_{\eta\eta} \approx 0 $. 
In this case 
the effective range expansion in Eq.~\ref{s3a} is not appropriate. 
A better parametrisation is obtained  with simply an expansion of 
$K_{\eta\eta}(E)$ 
instead of $1/K_{\eta\eta}(E)$. Actually the zero in the scattering length 
occurs in the subthreshold region. This solution resembles to some 
extent the case of elastic $\pi N $ scattering.
It should be added that those analyses that extract the $\eta N$
scattering length ($a$) from a many hadron reaction -- such as $pn\rightarrow d\eta$\cite{Speth},
 $\eta d \rightarrow \eta d$ \cite{Shev} or $ pd \rightarrow \  ^{3}He \eta$
 \cite{wil} -- are, in fact, discussing $a$ at an energy of 10 -- 20 MeV
{\em below} the threshold. As can be seen from  Fig.~\ref{fig1.}a) this
results a value of $T_{\eta\eta}$ that is substantially smaller than the
threshold value. Therefore, in those cases where an energy
extrapolation is not performed,  
one should not be surprised, if the values of $a$
quoted as being appropriate for such many hadron reactions are smaller
than from those reactions involving at most two hadrons -- as in the 
$K$-matrix description by the present authors.

Since the extension of the $K$-matrix model to three explicit channels
removes the need to choose a form for $A(phot)$, it is now possible to
check whether or not the unconventional solution GW21 in
Ref.~\cite{gw98} is still acceptable. Now the most striking difference in
the parameters  in Table III of Ref.~\cite{gw98} is seen
in the $B_{\eta \eta}$, which is 0.11 for GW11 and --0.83 for GW21.
There is a direct correspondence between $a$ and $B_{\eta \eta}$ -- the
more positive $B_{\eta \eta}$ gives the more positive $a$.
However, in spite of trying many different starting values in
the Minuit minimization program -- especially starts with large negative
values for the background term $B_{\eta \eta}$ -- the GW21-like solution
could not be found. This suggests that the energy dependent form of 
$A(Phot)$ in Eq.~\ref{A} is not a good assumption. However, this is not
the case. The extended $K$-matrix model now supplies the input for
checking both assumptions for $A(Phot)$ -- the energy independent form
and the one in  Eq.~\ref{A} -- and in both cases the forms are found to
be very nearly proportional to $T_{\gamma \eta}$ -- the differences
being very small. This now suggests that the appearance and
disappearance of the unconventional solution could be due to two almost
degenerate minima in the parameter space -- a feature that will rise
again later.

\begin{figure}[ht]
\includegraphics{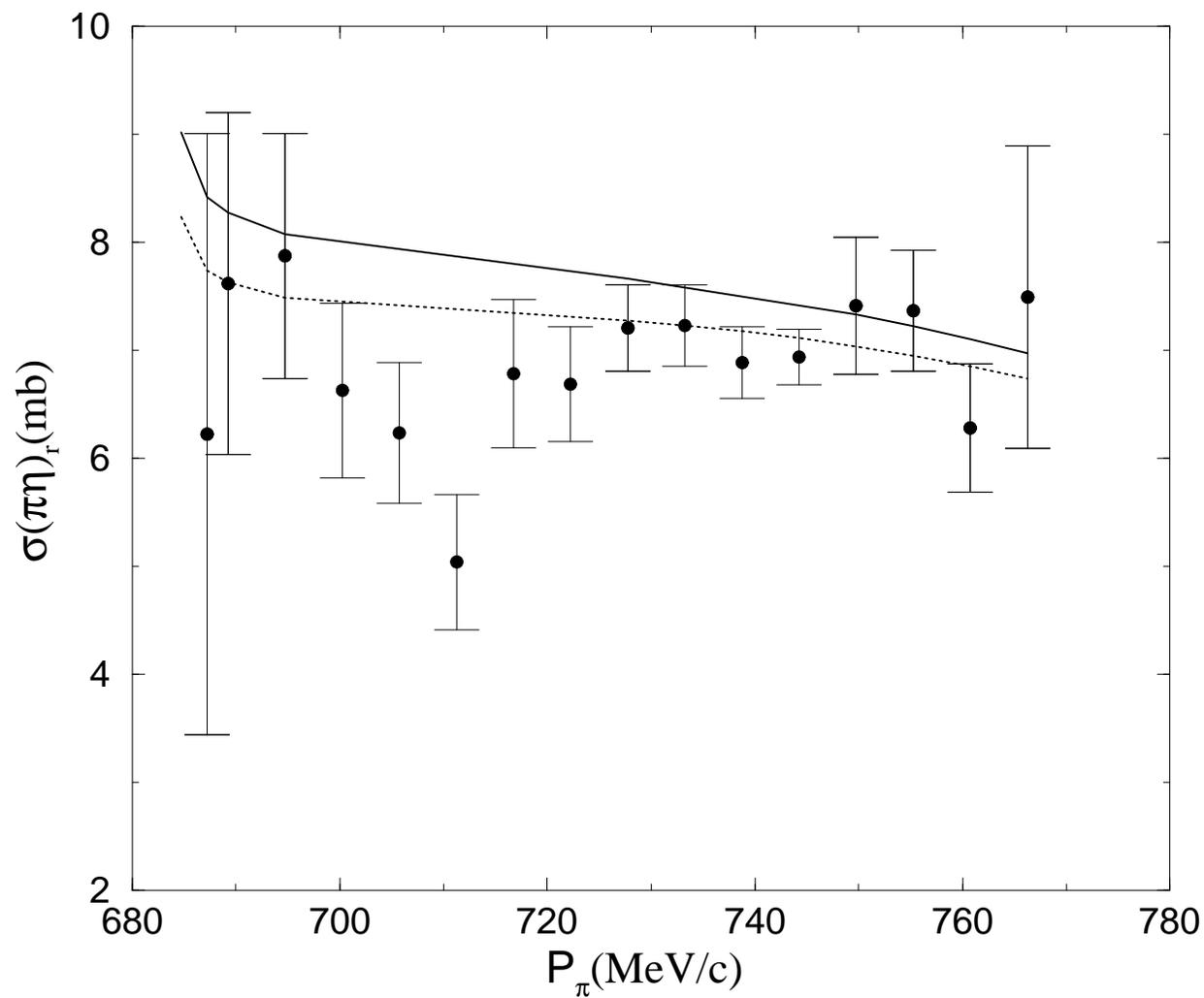}
\caption{The reduced $\pi N\rightarrow \eta N$ cross section 
$\sigma(\pi\eta)_r$ in mb as a function of the $\pi$-momentum in MeV/c
-- see Refs.\protect \cite{Tom} and \protect \cite{Igor}.
Solid line -- conventional solution(C), 
Dotted line -- unconventional(UC) solution.}
\label{fig2.}
\end{figure}

Since the publication of Ref.~\cite{gw99} there has become available more
experimental data: 

 a) Firstly, the GRAAL collaboration has extended 
the $\gamma p\rightarrow\eta p$ data of  Krusche et al. \cite{Krusche} 
from the range $1486 <E_{cm}< 1539$MeV upto about $E_{cm}=1700$MeV 
\cite{GRAAL}. This new data not only gives total cross sections 
but also differential cross sections, which do not concern us at this stage.
When the new
total cross sections, at the higher energies, are superimposed on Fig.~3
in Ref.~\cite{gw99} depicting the $K$-matrix extrapolations beyond
$E_{cm}=1539$MeV, it is found that solution A corresponding to
$a(fm)=0.87+i0.27$ is favored over solution B.
 
 b) Secondly, for the  $\pi  N\rightarrow \eta N$ reaction there is now
 new data~\cite{Tom} and also a reassessment by Strakovsky et al.~\cite{Igor}
 of the earlier review by Clajus and Nefkens~\cite{Nefkens}.  
 The main difference compared with the earlier set is that, 
in the reduced cross section 
\[\sigma(\pi\eta)_r=\sigma(\pi\eta)\frac{q_{\pi}}{q_{\eta}},\]
there is now  a distinct structure over the pion laboratory momentum range
$700<p_{\pi}<720$ MeV/c -- see Fig.~\ref{fig2.}.
 In the earlier review ~\cite{Nefkens}, $\sigma(\pi\eta)_r$ 
 was almost constant with a small monotonic decrease upto  about 
770 MeV/c. If the data points in the range $700<p_{\pi}<740$ MeV/c are 
ignored, then the Minuit fitting procedure always results in a unique
solution independent of the parameter starting values. This appears to
be a robust result. However, if the
points in this range are included, then the resultant effective range 
parameters depend on the starting values -- particularly on  the
background term $B_{\eta \eta}$. Broadly speaking, if  the starting value of
$B_{\eta \eta}$ is chosen to be positive, then the resulting solution
 always has $Re \ a \approx 1$fm and the resonance widths are conventional.
 However, for negative starting values of $B_{\eta \eta}$, 
 the unconventional solution emerges with $Re \ a\approx 0.2$fm and 
 resonance shapes very different to the Particle Data Group~\cite{PDG}.
The resulting $\chi^2$/dof are very similar for the two cases being
respectively 1.036 and 1.029  with 155 data points and 12 parameters.
 This clearly shows that, within the present $K$-matrix model, there are
two minima in the parameter space. However, as seen in Fig.~\ref{fig2.},
the reason why the unconventional solution emerges, as the data points in
the range $700<p_{\pi}<740$ MeV/c are incorporated into the data base, is
that the featureless curve for  $Re \ a\approx 0.2$fm is lower than
the equally featureless curve for the higher values of $Re \ a$. Even so,
neither solution is able to account for the structure now seen in the 
reduced cross section. This result can be interpreted in several ways:

a) In the $K$-matrix model, some effect is omitted in 
$\pi N\rightarrow \eta N$ channel in the present
form of the parametrisation. But it is difficult to image anything
that could have such a dramatic effect so near the threshold.

b) The results in Refs.~\cite{Tom,Igor} need to be checked, since if
this structure is indeed confirmed, it indicates the presence of some hitherto
unexpected mechanism not seen in the other channels.
 
In Table  II, the parameters corresponding to the conventional (C) and
unconventional(UC) solutions are compared with each other and the values
given by the Particle Data Group \cite{PDG}. There it is seen that the
two solutions are qualitatively the same as GW11 and GW21 in our
earlier work~\cite{gw98}. Again the distinguishing features between the
two fits are the values of the energy of the $K$-matrix pole and the 
corresponding widths [$E_0$ and $\Gamma(1535)$ in Table II].
Of course, this is not reflected in the  resonance position and width
defined as the Real and Imaginary  resonance energies from the $T$-matrix.
In fact, they are almost the same for GW11 and GW21 with the values
$E_P=$1514 versus 1509MeV and  $\Gamma/2=$90 versus 82MeV respectively. 
The difference between the two solutions is the non-Lorentian shape of the 
resonant $T$-matrix in the GW21 case. This is seen clearly in eq.(18) of
Ref.\cite{gw99}. 
and follows the same analysis that leads here to  Eqs.~\ref{s2},\ref{s3}.

Even though the overall $\chi^2$/dof are virtually the same for the two
solutions, the individual $\chi^2$/dp for each type of amplitude are
much  different -- as can be seen in Table III.

One of the authors (S.W.) wishes to acknowledge the hospitality of the
Helsinki Institute of  Physics, where part of this 
work was carried out. The authors also thank Drs.T. Morrison, 
D. Rebreyend, F.  Renard and I. Strakovsky  for useful 
correspondence concerning their unpublished data..

\newpage

{\bf Figure Captions}

\vspace{1.0cm}

Fig 1: The $\eta N\rightarrow \eta N$ amplitude  $T(\eta N)=T_{\eta
\eta}$ in fm as a function of the center-of-mass energy $E_{C.M.}$ in MeV.
Solid line for Re $T$ and Dotted line for Im $T$. 
a) is for the conventional(C) solution GW11 and 
b) is for the unconventional(UC) solution GW21.

\vspace{0.5cm}

Fig 2: The reduced $\pi N\rightarrow \eta N$ cross section
$\sigma(\pi\eta)_r$ in mb as a function of the $\pi$-momentum in MeV/c
-- see Refs.\protect \cite{Tom} and \protect \cite{Igor}. 
Solid line -- conventional solution(C), 
Dotted line -- unconventional(UC) solution.
\newpage

\begin{table}
\begin{center}
\caption{ A selection of $\eta N$- scattering lengths  $a$ appearing in
the literature. }
\vspace{0.5cm} 
\begin{tabular}{ld} 
Reaction or Method & Scattering Length(fm) \\ \hline
Isobar model  \protect\cite{ben}&0.25+i0.16\\
Isobar model\protect\cite{bha}&0.27+i0.22 \\
$pn\rightarrow d\eta$ \protect\cite{Speth}& $\leq $0.3 \\
Isobar model \protect\cite{sau}& 0.51+i0.21 \\
$ pd \rightarrow \  ^{3}He \eta$ \protect\cite{wil}&0.55(20)+i0.30 \\
Coupled $T$-matrices \protect\cite{aba96}& 0.621(40)+i0.306(34) \\   
Effective Lagrangian \protect\cite{kai}& 0.68+i0.24\\
Coupled $K$-matrices\protect\cite{gw97}& 0.75(4)+i0.27(3)\\ 
$\eta d \rightarrow \eta d$ \protect\cite{Shev}&$\geq $0.75\\
Coupled $K$-matrices\protect\cite{gw99}& 0.87+i0.27\\ 
Coupled $T$-matrices\protect\cite{bat}, \protect\cite{new}&0.886+i0.274 \\ 
Coupled $T$-matrices\protect\cite{ari}& 0.98+i0.37 \\
Coupled $K$-matrices\protect\cite{gw99}& 1.05+i0.27\\ 
\end{tabular}
\label{table1}
\end{center}
\end{table}

\begin{table}
\begin{center}
\caption{ The optimised parameters  defining the
$K$-matrices for: a) a conventional(C) solution and b) an unconventional(UC)
solution compared with the Particle Data Group \protect\cite{PDG}. }
\vspace{1cm}
\begin{tabular}{cccc|cccc}
&C &UC&PDG&&C& UC&PDG\\ \hline
$B_{\eta \eta}$&0.4225&--0.4769&&$\Gamma(1535, \  Total)$(MeV)&125.5&267.0&100--250\\
$B_{\pi \eta}$&--0.0376&--0.0282&&$\eta(1535, \ br)$&0.590&0.578&0.30--0.55\\
$E_0$(MeV)&1529.7&1580.5&1535(20)&$\pi(1535, \ br)$& 0.346&0.358&0.35--0.55\\
$E_1$(MeV)&1681.3&1681.0&1650(30)&$\Gamma(1650, \ Total)$(MeV)&165.6&
164.6&145-- 190\\
$B_{\gamma\eta}$&0.00223&0.00707&&$\pi(1650, \ br)$&0.732&0.731&0.55 -- 0.90\\
$B_{\gamma \pi}$&0.00250&0.00484&&&&&\\
$\gamma_{\gamma}$&0.000166&0.000240&&&&&\\
\end{tabular}
\label{table2}
\end{center}
\end{table}

\begin{table}
\begin{center}
\caption{Comparison of the $\chi^2$/dp for the separate amplitudes.
In the $\pi  N\rightarrow \pi  N$ and $\gamma N\rightarrow \pi N$ cases,
the two entries refer to the contributions from the Real and
Imaginary terms respectively.}
\vspace{1cm}
\begin{tabular}{cccccc}
Solution&$\pi  N\rightarrow \eta N$&$\pi  N\rightarrow \pi  N$&
$\gamma N\rightarrow \eta N$&$\gamma N\rightarrow \pi N$& Total\\
C       &0.981                     &1.281, 0.803              &
0.869                       &0.556, 1.403&1.036\\
UC&1.017&1.206, 0.808&0.893&0.538, 1.334& 1.029\\
\end{tabular}
\label{table3}
\end{center}
\end{table}

\newpage

\newpage

\end{document}